# Highly cited references in PLOS ONE and their in-text usage over time

Wolfgang Otto[1], Behnam Ghavimi[1], Philipp Mayr[1], Rajesh Piryani[2], Vivek Kumar Singh[3]

[1] *{firstname.lastname}@gesis.org*
GESIS - Leibniz Institute for the Social Sciences, 50667 Cologne (Germany)

[2] *rajesh.piryani@gmail.com*
Department of Computer Science, South Asian University, New Delhi (India)

[3] *vivekks12@gmail.com*
Department of Computer Science, Banaras Hindu University, Varanasi (India)

**Abstract**
In this article, we describe highly cited publications in a PLOS ONE full-text corpus. For these publications, we analyse the citation contexts concerning their position in the text and their age at the time of citing. By selecting the perspective of highly cited papers, we can distinguish them based on the context during citation even if we do not have any other information source or metrics. We describe the top cited references based on how, when and in which context they are cited. The focus of this study is on a time perspective to explain the nature of the reception of highly cited papers. We have found that these references are distinguishable by the IMRaD sections of their citation. And further, we can show that the section usage of highly cited papers is time-dependent: the longer the citation interval, the higher the probability that a reference is cited in a method section.

**Introduction**
Scientific publications are highly structured texts which incorporate specific properties related to their references. The accessibility of full-text publications has broadened up the possibilities for analysing citation behavior and the usage of referenced publications in bibliometrics. The objectives of this research-in-progress paper are highly cited PubMed papers in a corpus of the open access journal PLOS ONE[1] during the period between 2006 and 2017. As a highly cited PubMed paper, all papers (n=666) are taken into account which are cited in more than 100 PLOS ONE papers of our corpus. We call these highly cited papers *top-666*.

The main goal is to describe the highly cited papers based on extracted information from our corpus; this includes metadata of citing publications and citation contexts. So the information we gather is not from the full-text of the referenced publication, but the citing publications. On the paper level, we use the year of publication of the citing publication. On citation context level we use information about the sections based on the IMRaD scheme[2] which is "the most used standard of today's scientific discourse" (Sollaci & Pereira, 2004). As a second distinguishing feature, we examine the co-citation count on the citation context level. With the latter information, it is possible to investigate how a paper is perceived over time based on the citation interval. The citation interval is the time distance of the citing paper to the time of publication of the reference. With this information, we are able to describe the top cited references based on how, when and in which context they are cited. We use this information to describe the "citation history" of a specific cited object over time.

Having a deeper understanding and methodology for the usage of gathered information from citations based on full text is, on the one hand, helpful to understand the temporal citation patterns of references, on the other hand, the information can be used to build better tools for information retrieval systems for suggesting related research literature (e.g. Huang et al., 2015).

---

[1] https://journals.plos.org/plosone/
[2] i.e. Introduction, Method, Results, and Discussion

In our study we will address the following research questions:
- RQ 1: In which IMRaD section are the top-666 references cited?
- RQ 2: How is the proportion of concrete IMRaD sections evolving over time?
- RQ 3: Is the number of co-citations on citation context level declining when the citation interval is becoming longer?

**Related Work**

Citations are an important parameter of connectivity of related research works. A lot of studies have focused on analysing citations for different purposes ranging from assessment of the quality of an article to tracing the flow of ideas on a topic. Sugiyama et al. (2010) have suggested that there could be two kinds of citation analysis: (1) Citation Counts and (2) Citation Context Analysis. They argue that citation context analysis could be a better approach to determine the influence of a research article. Unlike simple counts, citation context analysis identifies the contextual relationship between citing research articles and referenced articles by applying various NLP and Machine Learning approaches (Hernández-Alvarez & Gómez, 2016). It processes the text of articles, particularly that portion where it cites another article. This is called citation context, i.e. the sentence where a specific reference is cited. Relatively few studies have been carried out on citation context analysis.

Some researchers have performed a sentiment analysis by incorporating the citation context with the subjectivity analysis of citations (Athar & Teufel, 2012; Abu-Jbara et al., 2013; Athar, 2014). In a recent work, Bertin et al. (2016) have used the linguistic patterns to analyse the citation context and its location in the IMRaD structure of a research article to determine the recognition of citer motivation (Teufel et al., 2006). Another work by Small (2018) has studied the phenomenon of highly cited research articles based on citation context. They have examined citation context for linguistic patterns which are associated with different types of referenced research articles. As types, they consider method and non-method publications.

Boyack et al. (2018) have analysed the in-text citation characteristics in the larger dataset of Elsevier full-text journal articles and PubMed Open Access Subset articles. They have identified that all fields such as references, sentences, in-text citation numbers per article have increased over time. They also found that highly cited publications are often cited only once per publication. An et al. (2017) have carried out a study to identify the characteristics of highly cited authors on the basis of citation location and contexts using NLP techniques. They have worked on the ACL Anthology dataset (Bird et al., 2008). Atanassova & Bertin (2016) investigated positions of in-text citation in IMRaD structure regarding the age of cited papers. Similar to Boyack et al. (2018) and Bertin et al. (2016), we search for patterns in the position and the linguistic context references are cited. But in opposite to them, we change the perspective from the citing publication to the referenced publication. Analogue to Small (2018), we are interested in retrieving information for specific highly cited references from their citation context. But unlike Small, our goal is not the classification of references based on a predefined schema. We want to provide a basis for a comprehensive analysis of the references based on the citation contexts. Additionally, we introduce time aspects in respect to citation interval to the analysis of highly cited references.

**Methods**

*Description of the Dataset*

Our research object is a corpus of citation contexts of highly cited references. Each context consists of the sentence in which the citation occurs. For all sentences, we add information about the citing publication and basic information (incl. publication year) of the referenced publication. This information originates from the reference sections of the citing publications. No additional information source is used. To create the dataset, we start with a corpus of 176,856 papers from PLOS ONE published between 2006 and 2017. While we are interested in citation contexts of highly cited papers we selected all references which are cited in more than 100 PLOS ONE articles. To be able to get additional metadata and keep the problem of deduplication away, we choose only references with PubMed ids in the reference part of the citing publications. An example of a citation context with related metadata is shown in Table 1. In total, the number of references cited in more than 100 publications is 666. 127 publications are discarded because of missing PubMed id. In the next step, we filter all citation contexts which do not reference one of the top-666 publications. Because not every publication in our corpus is referencing one of these publications, the number of citing papers used in our study shrinks down to 62,127.

*Corpus statistics*

The number of relevant citation contexts for our analysis is 173,630. This number reflects the fact, that only in 0.5 percent of the citation contexts (total: 31,746,769) at least one top reference is cited. These top-referenced publications are published between 1951 and 2015. Only 69 of them are published before 2010. We have to keep in mind that just for cited articles published after 2007 it is possible to examine statistics for short citation intervals. The distribution of the number of citation contexts per top-666 follows a power-law-like shape. The most cited reference is mentioned in 3,363 citation contexts. This reference is a method paper by Livak and Schmittgen[3] titled "Analysis of relative gene expression data using real-time quantitative PCR and the 2(-Delta Delta C(T)) method". The lowest number of citation contexts is 75, where the median reference is mentioned in 184 contexts.

**Table 1. Example of extracted data for one citation context.**

| | |
|---|---|
| DOI of citing paper | 10.1371/journal.pone.0013678 |
| Citation context | *"The molecular mechanisms of nuclear reprogramming are still unsolved although recent reports have shown that reprogramming of human somatic cells can be achieved in vitro by retroviral expression of four transcription factors creating induced pluripotent stem (iPS) cells, which are comparable to ES cells [1], [2], [3], [4]."* |
| Pubmed-ID of reference | 18035408 |
| Section title | Introduction |
| Pub. year of the citing paper | 2010 |
| Pub. year of the reference | 2007 |
| Citation interval | 3 years (2007-2010) |
| No. of co-citations | 4 |

---

[3] This paper is also the top 1 cited paper in a similar study by Small (2018).

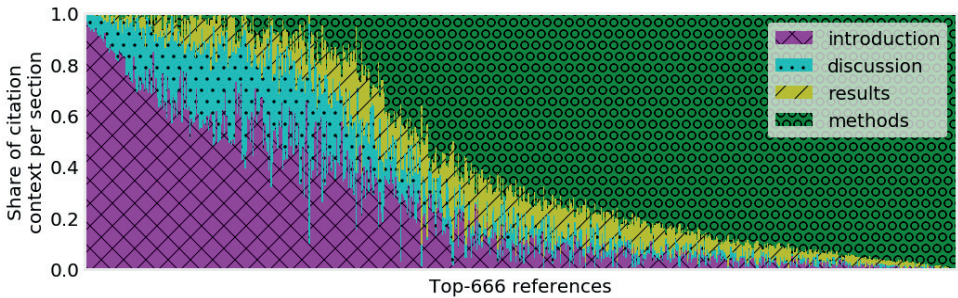
**Figure 1. Share of section where all top-666 references are cited.**

**Results**

*Distribution over sections*

The first question we want to answer with the data from our corpus for the top-666 references is: In which section are the top-666 cited (RQ 1)? Small (2018) used the share of citation contexts located in the method section of a citing publication to predict what type of reference it is. The usefulness of this feature to predict the type of the referenced publication provides a hint, that their usage in a specific section could distinguish highly cited references. Figure 1 shows for each of our top-666 references on the x-axis the proportion of sections in which the citation contexts were found. For more than half of our references the most used section for citing is the method section. But on the other hand, there is a larger group of references which are mainly used in introduction and discussion. We tried to correlate the section usage with the mean citation interval of the top-666. This did not show a significant result. In general, this means, that the section usage of a reference is not dependent on the age of a reference while citing.

*Sections of citation contexts over time*

To be able to delineate the influence of age of referenced literature and the type of the section in the citing papers we measure the citation interval (RQ 2). The citation interval is the time distance between the publication date of the citing paper and the publication date of the reference paper. We accumulate all citation contexts in groups for each possible citation interval to the corresponding reference. This grouping results in a minimum of zero years to a maximum of 65 years. Of course, the amount of citation contexts in each group varies. Eighty percent of the citation intervals are between 5 to 20 years. For citation intervals larger than 25 years, we have not much account on absolute numbers of citation contexts as well as number of different referenced objects. For each group, we calculated the share of sections and visualized the results in Figure 2.

The result of this analysis is that a reference with a high citation interval, i.e. an older publication at the time of citing, is likely to be used in a method part. The second result is that there is a change for the most frequently used section based on the citation interval at the beginning of a reference lifetime (0-3 years). In the first two years, a top-cited publication is more likely to be used in the introduction part, later it is more often used in the method part. This usage is a hint that the function of the citation is changing over time and needs further examination in future work.

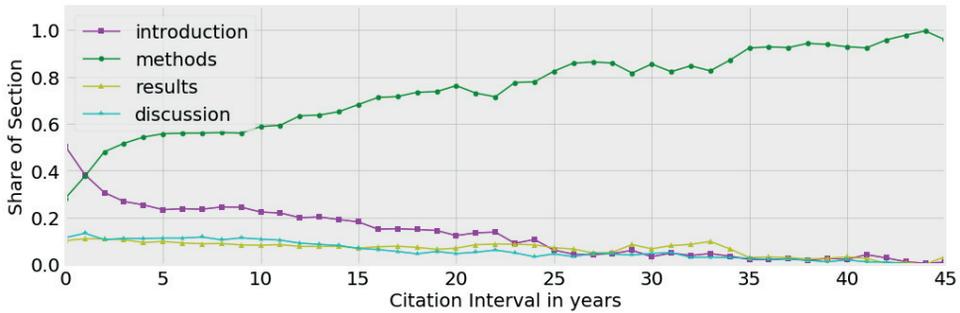

**Figure 2. The proportion of IMRaD sections over time.**

*Number of co-citations on citation context level*

The third introspection (RQ 3) of the usage of highly cited references concerns their co-citation on the citation context level. For each relevant context based on citing a top-666 reference we count the number of all references. This approach is defined by Liu & Chen (2012) as sentence-level co-citation. The fact that a publication is referenced alone in one citation context could be a sign that the reference stands on its own and there seems to be no need of citing similar publications. The larger the citation interval is, the higher could be the probability, that there is a low number of co-cited publications in the same sentence.

To investigate the number of co-citations depending on the citation interval, we binned all citation contexts for each citation interval (0-40 years) separately. For each year we have calculated the mean value from the number of quotations per citation context. The mean value is between 1 and 2. Figure 3 (a) shows that the mean value is declining. We calculated the coefficients of a linear regression describing this correlation which is *-0.54* and stated to describe a significant relation. But probably we have to include the knowledge that the higher the citation interval, the greater the probability that the citation context is in a method part of a paper. This fact raises the question whether this phenomenon is also able to describe the declared co-citation time effect.

Here we can come up with a derived research question: Is the high proportion of "method"-citation contexts within longer citation intervals explaining the change of co-citation mean? To answer this question, we divide the citation context into four groups based on location in one of the sections *Introduction, Methods, Results, and Discussion*. After this grouping, we applied the same procedure as for the first calculation. Figure 3 (b) reflects the mean co-citation number in the context given a citation interval for different sections. For all the curves we figured out that there are no significant relations between citation interval and the number of co-citations. We combine the results of this non-significance with the knowledge that the proportion of the citation context of the method increases due to the length of the citation interval (Figure 2). Then, the fact that the mean number of co-citations in the method-contexts is the lowest (Figure 3 (b)) explains the decreasing curve of Figure 3 (a).

**Future Work**

Future studies aim to replicate results by concerning a larger data corpus, for example by considering the whole PubMed corpus. Also, future research should consider the potential effects of various citation intervals more carefully. For example, analysis of word usage (e.g., adjective, verb, and noun) in citation contexts of highly cited papers and changes over time might be addressed. Besides, the categorization of the cited works into different types or disciplines by their citation contexts could be an essential field for future research.

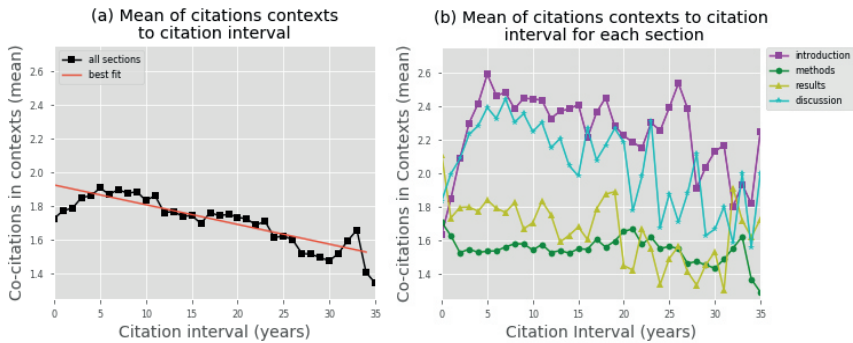

Figure 3. The curves describe the decrease of the mean number of co-citation in contexts for longer citation intervals. (a) Describes the overall decline incl. the best-fit line.
(b) Reflects the means for each section individually.

**Acknowledgments**

The authors acknowledge the enabling support provided by the Indo-German Joint Research Project titled 'Design of a Sciento-text Computational Framework for Retrieval and Contextual Recommendations of High-Quality Scholarly Articles' (Grant No. DST/INT/FRG/DAAD/P-28/2017) for this work.

**References**

Abu-Jbara, A., Ezra, J., & Radev, D. (2013). Purpose and polarity of citation: Towards nlp-based bibliometrics. In Proceedings of ACL 2013 (pp. 596-606).

An, J., et al. (2017). Exploring characteristics of highly cited authors according to citation location and content. JASIST, 68(8), 1975-1988.

Atanassova, I., & Bertin, M. (2016). Temporal properties of recurring in-text references. D-lib Magazine, 22(9/10).

Athar, A., & Teufel, S. (2012). Context-enhanced citation sentiment detection. In Proceedings of ACL 2012 (pp. 597-601). ACL.

Athar, A. (2014). Sentiment analysis of scientific citations (No. UCAM-CL-TR-856). University of Cambridge, Computer Laboratory.

Bertin, M., et al. (2016). The linguistic patterns and rhetorical structure of citation context: an approach using n-grams. Scientometrics, 109(3), 1417-1434.

Bird, S., et al. (2008) The ACL Anthology Reference Corpus: A reference dataset for bibliographic research in computational linguistics. In Proceedings of LREC 2008, (pp. 1–5).

Boyack, K. W., et al. (2018). Characterizing in-text citations in scientific articles: A large-scale analysis. Journal of Informetrics, 12(1), 59-73.

Hernández-Alvarez, M., & Gómez, J. M. (2016). Survey about citation context analysis: Tasks, techniques, and resources. Natural Language Engineering, 22(3), 327-349.

Huang, W., et al. (2015). A Neural Probabilistic Model for Context Based Citation Recommendation. In AAAI (pp. 2404-2410).

Liu, S & Chen C. (2012). The proximity of co-citation. Scientometrics, 91(2), 495-511.

Small, H. (2018). Characterizing highly cited method and non-method papers using citation contexts: The role of uncertainty, Journal of Informetrics, 12(2), 461-480.

Sollaci, L. B., & Pereira, M. G. (2004). The introduction, methods, results, and discussion (IMRAD) structure: a fifty-year survey. Journal of the medical library association, 92(3), 364.

Sugiyama, K., et al. (2010). Identifying citing sentences in research papers using supervised learning. 2010 International Conference on Information Retrieval & Knowledge Management (CAMP).

Teufel, S. et al. (2006). Automatic classification of citation function. Proceedings of EMNLP 2006.